# Direct Observation of Resonant Scattering Phase Shifts and their Energy Dependence

Stephen D. Gensemer,[†] Ross B. Martin-Wells, Aaron W. Bennett, and Kurt Gibble
Department of Physics, The Pennsylvania State University, University Park, PA 16802, USA

We scan the collision energy of two clouds of cesium atoms between 12 and 50 μK in atomic fountain clock. By directly detecting the difference of s-wave scattering phase shifts, we observe a rapid variation of a scattering phase shift through a series of Feshbach resonances. At the energies we use, resonances that overlap at threshold become resolved. Our statistical phase uncertainty of 8 mrad can be improved in future precision measurements of Feshbach resonances to accurately determine the Cs-Cs interactions, which may provide stringent limits on the time variation of fundamental constants.
PACS: 34.50.Cx, ρ06.30.Ft

Feshbach scattering resonances occur when the continuum state of two colliding particles couples to a bound state (Fig. 1) [1]. Feshbach resonances have found wide applicability in dilute, ultracold, atomic and molecular gases because they provide an accessible control of the inter-particle interactions [2-6]. Feshbach's elegant treatment of scattering resonances showed that scattering phase shifts, and hence cross sections, change rapidly as the collision energy tunes through resonance. The rapid phase change is a general feature of resonance phenomena and the resonant energy dependence of cross sections has been observed in a variety of experiments, including neutron and electron scattering and photodetachment [7-9]. In ultracold gases, so far magnetic fields have been used to tune resonances to threshold, changing the energy of the bound state by vertically translating the grey potential in Fig. 1, instead of tuning the collision energy [3-11]. Here, we scan the collision energy between two ultracold clouds of cesium atoms in an atomic clock and directly observe the scattering phase shift [12] through a series of scattering resonances. Increasing the collision energy allows us to resolve resonances that overlap at threshold. Precise measurements of scattering phase shifts through a resonance will very accurately determine the resonance position, giving a highly precise determination of the atomic interactions [5] and a potential route to stringent limits on the time variation of fundamental constants [13,14].

We directly measure scattering phase shifts by preparing cesium atoms in coherent superpositions of the two clock states and detecting the phase shift of these coherences after the clock atoms scatter off atoms prepared in a pure 'target' state (Fig. 2 inset) [12]. When the $|F=3,m_F=0\rangle$ ($|40\rangle$) clock state scatters off the target atoms, it acquires a scattering phase shift $\delta_3(\delta_4)$. The phase of the clock coherence, the superposition of $|30\rangle$ and $|40\rangle$, precesses as hands on a clock. The scattering causes the phase of the coherence to jump by the difference of the scattering phase shifts, $\Phi=\delta_4-\delta_3$ [12], represented by the time difference between the ring of scattered clocks and the unscattered clock in the Fig. 2 inset. We directly detect $\Phi$ as a phase shift of the clock's Ramsey fringes as in Fig. 2(b). To be sensitive to the s-wave phase shifts, we detect atoms that scatter near 90°, using the Doppler shift of a stimulated-Raman transition [12,15]. By also measuring the cross sections for each clock state on $|32\rangle$, we isolate the scattering resonances to only the $|32\rangle \otimes |40\rangle$ channel, and exclude the $|32\rangle \otimes |30\rangle$ channel.

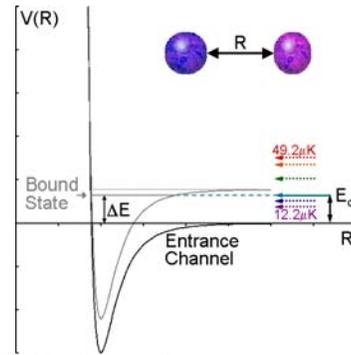

Fig. 1. A Feshbach scattering resonance occurs when the collision energy $E_c$ (aqua solid arrow) has the same energy as a bound state of the two particles. We scan the collision energy from 12 to 49 μK (dashed arrows) and observe a series of Cs Cs-scattering resonances. The schematic grey potential energy surface, as a function of internuclear distance R, can also be shifted vertically by changing the magnetic field B, which instead shifts the energy of the bound state $\Delta E$. Throughout the paper we use energy units of $E_c/k_B$.

In our fountain clock, we launch two clouds of cesium atoms with short time delays, 6.5-11 ms, to give collision energies between 12 and 50 μK [12,16]. A series of microwave pulses in three microwave cavities state prepares the atoms in the first cloud (C1) in a desired $|3,m_F=1,2,3\rangle$ target state and the second cloud (C2) in $|41\rangle$. A two-photon stimulated Raman transition transfers a 270 nK wide slice of the C2 vertical velocity distribution from $|41\rangle$ to $|30\rangle$. Interleaved with these transitions are clearing laser pulses tuned to the $6s_{½},F=4 \rightarrow 6p_{3/2},F'=5'$ and $2\rightarrow3'$ transitions that remove unselected atoms. The microwave clock cavity then prepares C2 atoms in a coherent superposition of $|30\rangle$ and $|40\rangle$ and the clouds collide near the apogee. A small fraction of the atoms in the clouds scatter, forming an expanding spherical shell of atoms, whose coherence is shifted by the difference of their s-wave scattering phase shifts. When the scattered shell is centered in the clock cavity, a second π/2 clock pulse converts the phase of the scattered atoms' coherence into a population difference between $|30\rangle$ and $|40\rangle$. A clearing pulse then removes the C1 target atoms and the atoms in $|30\rangle$. To detect the scattered atom's Ramsey fringe, we velocity-selectively transfer atoms that scatter near 90° from $|40\rangle$ to $|30\rangle$. After clearing and optical pumping pulses remove F=4 atoms, a laser beam tuned to F=4 → F'=5 excites the atoms and we collect the fluorescence with a lens and photodiode. To detect the phase shift of the scattered atoms, we use the phase of the unscattered atoms as a reference (Fig. 2(B)) and, to measure and remove backgrounds, we use a pump-probe technique where we inhibit the selection of either C1, C2, or both clouds [12].

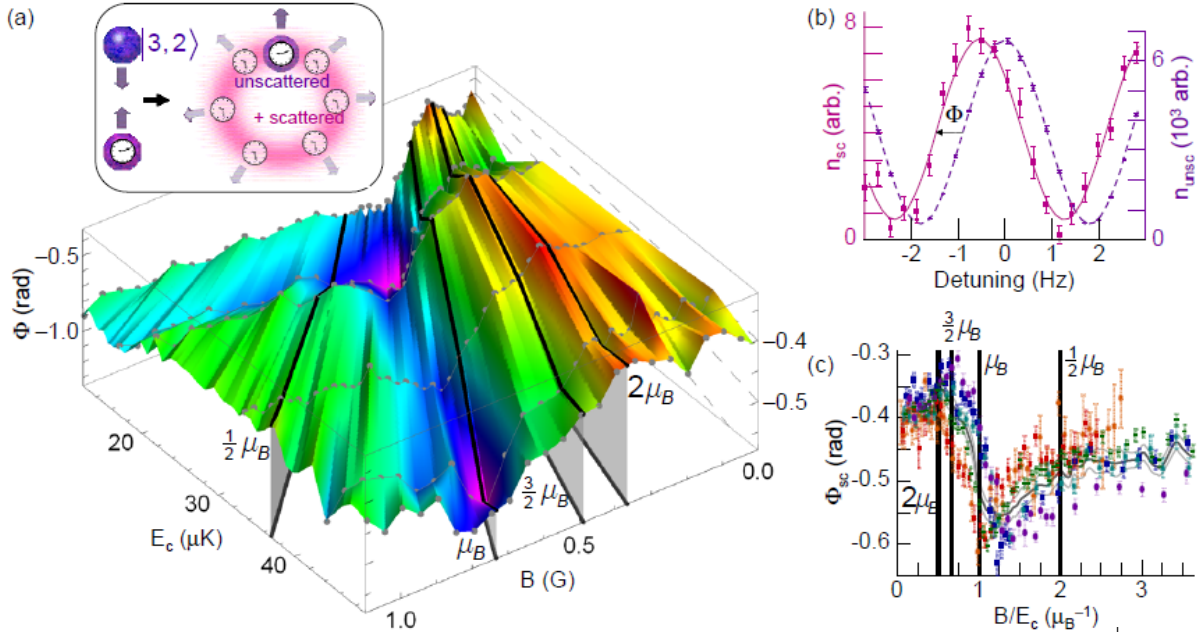

Fig. 2. (Inset) Clock atoms in a coherent superposition of two internal clock states scatter off of targets atoms (in $|32\rangle$), giving each clock state a quantum scattering phase shift. The phase of the clock coherence is represented by the time of the scattered and unscattered clocks. For scattered atoms, the clock phase is shifted by the difference of the quantum scattering phase shifts $\Phi$ of the two clock states. (a) Resonances of quantum scattering phase shifts versus collision energy $E_c$ and magnetic field B. The measured phase differences $\Phi$ (grey dots) are the difference of s-wave phase shifts for the cesium clock states scattering off cesium atoms prepared in $|32\rangle$. The black lines illustrate the allowed differences of magnetic moments for an l=0 halo molecular state. Features at $\mu_B$ and $3/2\mu_B$ suggest the bound-states are resonant near $E_c$=0. For clarity, the vertical scale varies linearly with $E_c$. (b) Atomic clock transition probability with two π/2 pulses for scattered (solid) and unscattered (dashed) atoms at $E_c$=12.2 μK and B=0.3 G. The phase shift $\Phi$ is plotted in (a). Each point represents the average of 18 fountain launch measurements, plus background measurements [16]. (c) $\Phi$ from (a) versus B, scaled by $E_c$, yields essentially the same shape for all $E_c$, as expected for the peculiar case of bound states near threshold. Here the amplitude of $\Phi$ is scaled by the spread of collision energies δE as discussed in the text. The black vertical lines represent the same magnetic moments in (a) and the data point colors are from Fig. 1. The weighted average of $\Phi_{sc}$ versus B gives an approximate reference shape for all energies (grey curves).

Fig. 2 shows a large and rapid variation of the scattering phase shifts $\Phi$ as a function of collision energy $E_c$ and magnetic field B for target atoms prepared in $|32\rangle$. Feshbach resonances generally tune with both the collision energy $E_c$ and magnetic field B, with a slope in the $E_c$ - B plane that is given by the difference of the magnetic moments of the coupled bound state and the incident channel. The black lines in Fig. 2 show the slopes for magnetic moments of ½, 1, 3/2, and 2 Bohr magnetons, $\mu_B$. Strikingly, all of the features we observe for $|32\rangle$ target atoms, as well as $|33\rangle$ and $|31\rangle$ targets below, follow one of these slopes with intercepts that are near the origin of the $E_c$ - B plane. For a single narrow Feshbach resonance, the observed width would be proportional to the spread of collision energies, roughly proportional to $E_c^{1/2}$. Instead, the features in Fig. 2 follow the black lines, scaling linearly with energy, as expected if there were two (or more) Feshbach resonances, with different magnetic moments. The small intercepts of the features suggest that the bound states have small binding energies (at B=0), lying very close to threshold, $E_c$=0. Further, if the slopes are multiples of $\mu_B/2$, it is likely that the bound states are halo molecular states, which have magnetic moments that are essentially the same as those for free atoms. For s-wave Feshbach resonances, the halo states asymptotically correspond to $|31\rangle \otimes |41\rangle$, $|30\rangle \otimes |42\rangle$, $|3,-1\rangle \otimes |43\rangle$, and $|3,-2\rangle \otimes |44\rangle$ at large internuclear separations. The magnetic moments for these halo states near threshold give resonance positions near the black lines in Fig. 2. We note that coupling to these states conserves the total z angular momentum, the total molecular $M_F$, as required for an s-wave resonance. If the Feshbach resonances had an angular momentum higher than s-wave, other total $M_F$'s would be allowed, with corresponding differences of magnetic moments. We have not observed these for target atoms in $|32\rangle$, or $|31\rangle$ or $|33\rangle$.

The shape of the resonant phase shifts in Fig. 2(a) versus magnetic field is remarkably similar for all of our collision energies $E_c$. If the corresponding bound states for all the Feshbach resonances have the same binding energy $\Delta E \approx 0$, scaling the magnetic field B as $B/(E_c+\Delta E)$ should align all of the resonances, as in Fig. 2(c). As for any general resonance, the scattering phase shift goes through $\pi$ as we sweep across a resonance. However, if each resonance is narrower than our rms spread of collision energies $\delta E$, the peak-to-peak phase excursions will be smaller, with an amplitude proportional to $\delta E^{-1}$. We prepare the atoms in each cloud with a mean temperature of 590 nK, which gives $\delta E$=3.2 $\mu$K at $E_c$ =12.2 $\mu$K and a larger $\delta E$=8.9 $\mu$K at 49.2 $\mu$K, proportional to $E_c^{1/2}$ [16]. Thus, the observed peak-to-peak phase excursions are smaller for large $E_c$. With this phase and $B/E_c$ scaling, the magnetic field dependence through the series of the Feshbach resonances in Fig. 2(c) has nearly the same form over our range of energies. Averaging this scaled data gives a function (grey curve) to compare to the measured phase versus $E_c$ and B. Interestingly, the grey curve and the data for each $E_c$ show the statistical significance of a nearly flat region between 3/2 and 1$\mu_B$; the phase steeply decreases, flattens, and then continues steeply down again for all of the collision energies, which can occur if there are resonances near 3/2 and 1$\mu_B$. We speculate that a Feshbach resonance should also exist near ½$\mu_B$. If it does, it is apparently sufficiently narrow that we do not resolve it for $|32\rangle$ targets, but the bound state coupling is evidently larger for $|33\rangle$ target atoms as we observe this resonance below.

Figure 3(a) depicts the bound state energies versus magnetic field (black lines), for multiples of $\mu_B/2$ with $\Delta E$ =0. The intersections of these bound state energies with our six chosen collision energies (horizontal lines) represent the positions of potential Feshbach resonances. In Fig. 3(c) the scattering phase shift difference for the $|32\rangle$ target atoms is shown for each $E_c$ versus B, along with the grey curve, the average of the scaled data in Fig. 2(c). While this curve describes the data quite well, there are deviations that vary systematically with energy. These deviations could be due to the multiple bound states having significantly different binding energies $\Delta E$. In addition, biases in the effective collision energy can occur since the $|32\rangle \otimes |40\rangle$ and $|32\rangle \otimes |30\rangle$ scattering cross sections are energy dependent. Both contribute to an energy bias of the scattered atoms that we detect. Full coupled-channels calculations yield both the phase and cross-section variations and a comparison of these with such measurements can precisely test and improve our knowledge of the Cs-Cs molecular interactions. We find that all the resonances, when extrapolated to B=0, are consistent with binding energies between 0 and 10 $\mu$K.

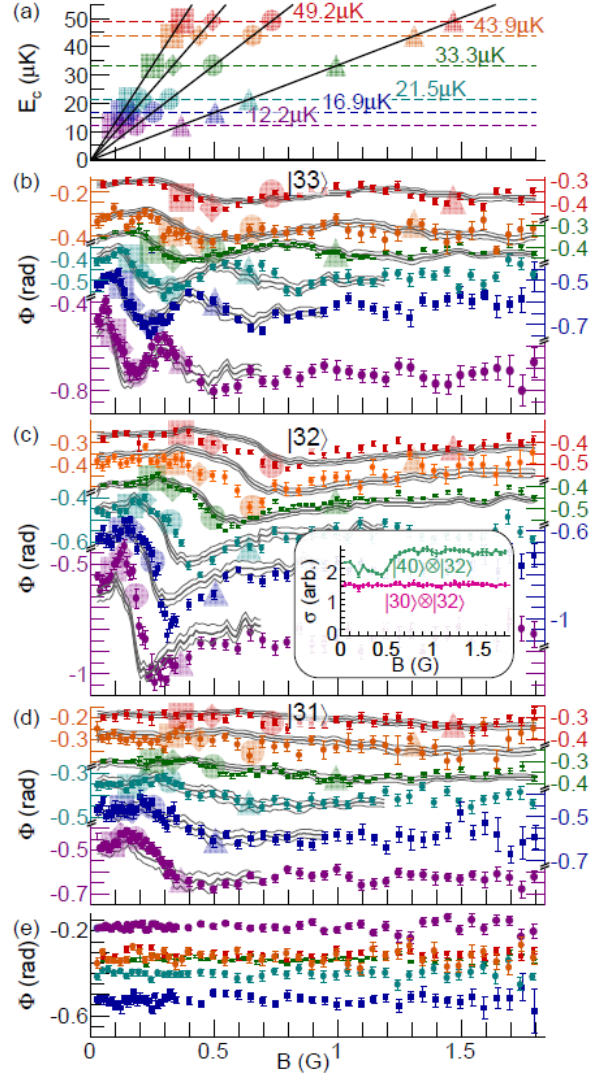

Fig. 3. (a) Positions of Feshbach scattering resonances versus $E_c$ and B, denoted by the large symbols (squares, diamonds, circles, triangles), for molecular bound states that are resonant at threshold, with magnetic moments that are different from those for bare atoms by (½, 1, 3/2, 2) $\mu_B$. The experimental $E_c$'s are depicted by the dashed horizontal lines. (b-e) Measured difference of s-wave scattering phase shifts for 6 collision energies $E_c$ versus magnetic field B, for target atoms in $|33\rangle$, $|32\rangle$, $|31\rangle$, and $|3,-3\rangle$. As in Fig. 2(a), the scattering resonances shift with $E_c$ and B. In (b-d) at all energies, scattering resonances occur near the symbols from (a) (see text). (c) Inset: The scattering cross section of $|32\rangle$ on $|40\rangle$ (magenta) shows scattering resonances, whereas $|32\rangle$ on $|30\rangle$ does not. In (e), note that the target for 12.2 $\mu$K is $|44\rangle$.

This is consistent with a previous observation of frequency shift cross sections at threshold, which appeared to indicate the existence of one or two overlapping Feshbach resonances of states with binding energies of $\Delta E$ =0 < 0.5 $\mu$K at B=0 [17,18].

Figures 3(b) and (d) show the measured phase shift differences for target atoms in $|33\rangle$ and $|31\rangle$. The $|33\rangle$ and $|31\rangle$ scattering shows features near the magnetic

moments of the same halo molecular states as $|32\rangle$. Here, $|33\rangle$ scattering has a clear resonance that asymptotically corresponds to $\approx\mu_B/2$ for the $|32\rangle \otimes |41\rangle$ exit channel.

Fig. 3(e) shows the phase difference for a $|3,-3\rangle$ target, which has no obvious resonances. We use this channel to measure small magnetic field gradients throughout the atomic trajectories in our fountain clock [16], which make a clock precision of 8 mrad challenging at high magnetic fields. While the clock transition, $|30\rangle \rightarrow |40\rangle$, has no linear Zeeman shift, it has a quadratic Zeeman shift of 427 Hz/G$^2$. At B=1.8 G, the quadratic Zeeman shift is 1.4 kHz. With a typical Ramsey fringe linewidth of 4.4 Hz for 33.3 µK, an 8 mrad phase uncertainty corresponds to a frequency uncertainty of 11 mHz, and a magnetic field uncertainty of 7 µG. Since we sweep over a wide range of magnetic fields, our clock has no magnetic shielding. To achieve this precision, a flux-gate magnetometer and control system actively stabilizes the vertical field, reducing the background fluctuations of approximately 3 mG. We use the unscattered clock atoms [Fig. 2 inset and (b)] to measure the quadratic Zeeman shift along the fountain trajectory. Because the s-wave halo of scattered atoms follows a different trajectory than the unscattered atoms, magnetic field variations in our clock produce a phase shift of order 100 mrad for B > 1G. We therefore interleave measurements of $|44\rangle$ or $|3,-3\rangle$ target atoms, prepared with a non-adiabatic magnetic field reversal (Fig. 3(e)). Since the phase shifts for these states show no scattering resonances, we use $|3,-3\rangle$ and $|44\rangle$ targets to measure and correct the magnetic field variations for the scattered halo for other target states [16].

We note that scattering phase shifts have sharp steps as inelastic scattering channels close. While scattering phase shifts wrap through π at Feshbach resonances, phase steps at inelastic thresholds do not, and are often small [19,20]. The phase excursions we observe are large, even after thermally averaging. Therefore, these are almost certainly Feshbach resonances, along with smaller contributions from inelastic thresholds. The peak-to-peak phase excursions we see are consistent with Feshbach widths of order of several mG. If the widths were significantly wider (narrower), our spread of collision energies would yield larger (smaller) peak-to-peak phase excursions.

The spread of collision energies for two colliding clouds is significantly larger than the temperatures of each cloud for high collision energies. The spread is $\delta E=[E_0(T_1+T_2)+3/8(T_1+T_2)^2]^{1/2}$, where $E_0$ is the collision energy due to the relative velocities of the two clouds (T=0 limit) and $T_1$ and $T_2$ are the temperatures of each cloud [16]. In our fountain clock, the atoms are cooled using degenerate sideband cooling in a moving-frame optical lattice [21]. The first cloud's temperature ($\approx$790 nK) is higher than the second's ($\approx$390 nK) since the first cloud, multiply loaded in a double-MOT [12], has more atoms than the second, which is launched directly from a vapor-cell MOT.

In conclusion, we directly observe a rapid variation of s-wave scattering phase shifts as we scan the collision energy through a series of Feshbach resonances. Reducing the spread of collision energies could yield even more precise measurements of the phase shifts and the positions of the Feshbach resonances. To resolve these overlapping resonances, the collision energy has to be of order 10 µK. Currently, our state-preparation velocity selects the second cloud to 270 nK and this velocity selection can be narrower. Adding a narrow velocity selection of the first cloud will help even more. Since only the velocity spread along the collision axis contributes significantly to the spread of collision energies, cooling in one dimension is sufficient [22]. Alternatively, a sample can be evaporatively cooled, split, and then accelerated to collide [23,24]. Higher energy resolution and a thorough evaluation of systematic errors will lead to a highly precise determination of the cesium interactions [5,25,26]. In turn, these can place stringent limits on the time variations of underlying fundamental forces [13,14].

We acknowledge discussions with S. Kokkelmans and E. Tiesinga and financial support from the NSF and Penn State.

†Present address: Centre for Engineered Quantum Systems, School of Physics, The University of Sydney, NSW Australia.

**Supplemental Material:**

**Experiment Details:**

Figures S1 and S2 show a schematic of our atomic fountain clock and a typical state preparation and detection sequence. We launch two clouds of atoms with time delays between 6.5-11 ms to give collision energies between 12 and 50 µK. A moving-frame optical lattice in the UHV MOT chamber cools the atoms to 0.3-1 µK via degenerate side-band cooling [12,21] and simultaneously optically pumps ≈75-85% of the atoms into $|33\rangle$. A sequence of microwave pulses, in three short microwave "state-preparation cavities" just above the UHV MOT, prepares the atoms in both clouds in the desired $|F,m_F\rangle$ state. To purify the spin composition of the colliding clouds, microwave pulses in the first selection cavity (SC1) transfer both clouds from $|33\rangle$ to $|43\rangle$ in steps A and D in Fig. S2. To prepare cloud 1 (C1) in $|31\rangle$, pulses B and C transfer the atoms from $|43\rangle$ to $|32\rangle$ to $|42\rangle$ with a continuous frequency sweep. To prepare C1 in $|33\rangle$ or $|32\rangle$, B and C are inhibited. After these microwave pulses, atoms remaining in F=3 are cleared with a laser pulse resonant with the $6s_{1/2}$,F=3 → $6p_{3/2}$,F'=2 cycling transition (e). Finally, C1 atoms in $|42\rangle$ are transferred to $|31\rangle$ (F) and cloud 2 (C2) from $|43\rangle$ to $|32\rangle$. For $|33\rangle$ or $|32\rangle$ targets, F instead drives the atoms in C1 from $|43\rangle$ to the target state In step (G) atoms remaining in F=4 are cleared with a laser pulse resonant with $6s_{1/2}$,F=4 → $6p_{3/2}$,F'=5 (G). C2 atoms are subsequently transferred from $|32\rangle$ to $|41\rangle$ in the third state-preparation cavity (H). Finally, a 270 nK slice of C2 atoms are velocity-selected with a two-photon stimulated Raman transition from $|41\rangle$ to $|30\rangle$. This, along with an F=4 clearing pulse at I, removes the wings of the velocity distribution in the vertical (z) direction for C2 to reduce the eventual detected background of unscattered atoms.

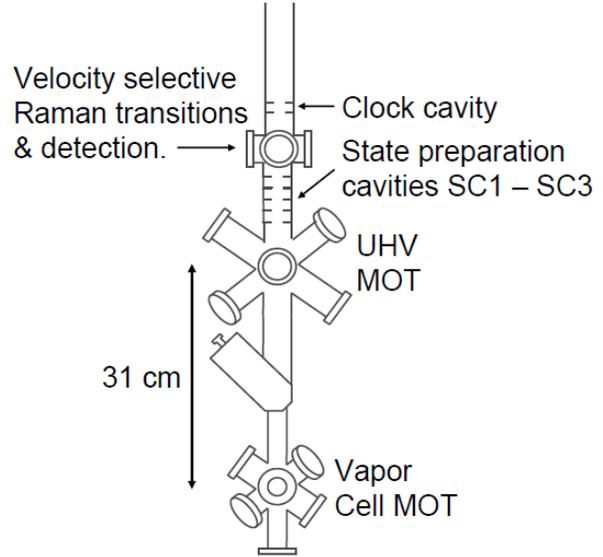

Fig. S1. Cloud 1 is repeatedly loaded from the Vapor Cell MOT into the UHV MOT. Cloud 2 is loaded and launched from the Vapor Cell MOT. Microwave state-preparation cavities SC1-SC3 prepare the atoms in the desired internal state. The apogee of the atoms is near the top of the microwave clock cavity. The schematic is to scale.

This series of microwave, stimulated Raman, and clearing pulses prepare the atoms in C1 and C2 in the desired states. A typical purity is more than 99%, and at least 95% for the worst case of $E_c$=12.2 µK. Each microwave state-preparation pulse is a frequency sweep and transfers ≈95% of the atoms. We avoid an overlap of the frequency sweeps with neighboring transitions to ensure that <1% of the atoms will be transferred out of or into adjacent $m_F$ states. To prepare C1 in $|3,-3\rangle$, we rapidly reverse the magnetic field to drive a Majorana transition from $|33\rangle$ to $|3,-3\rangle$, before C2 is captured in the molasses and lattice, before step A. Because the clearing beam pulses propagate vertically through the fountain, in the directions indicated, they illuminate both clouds so the state preparation sequence has to ensure that neither cloud is in the hyperfine state being cleared (upper or lower row in Fig. S2). The beams for the velocity-selective Raman transition originate from two phase-locked diode lasers that are detuned ≈2 GHz to the red of the $6s_{1/2}$,F=4 → $6p_{3/2}$ and $6s_{1/2}$,F=3 → $6p_{3/2}$ transitions. The beams subtend 6.6° such that $\mathbf{k}_1 - \mathbf{k}_2$ is vertical.

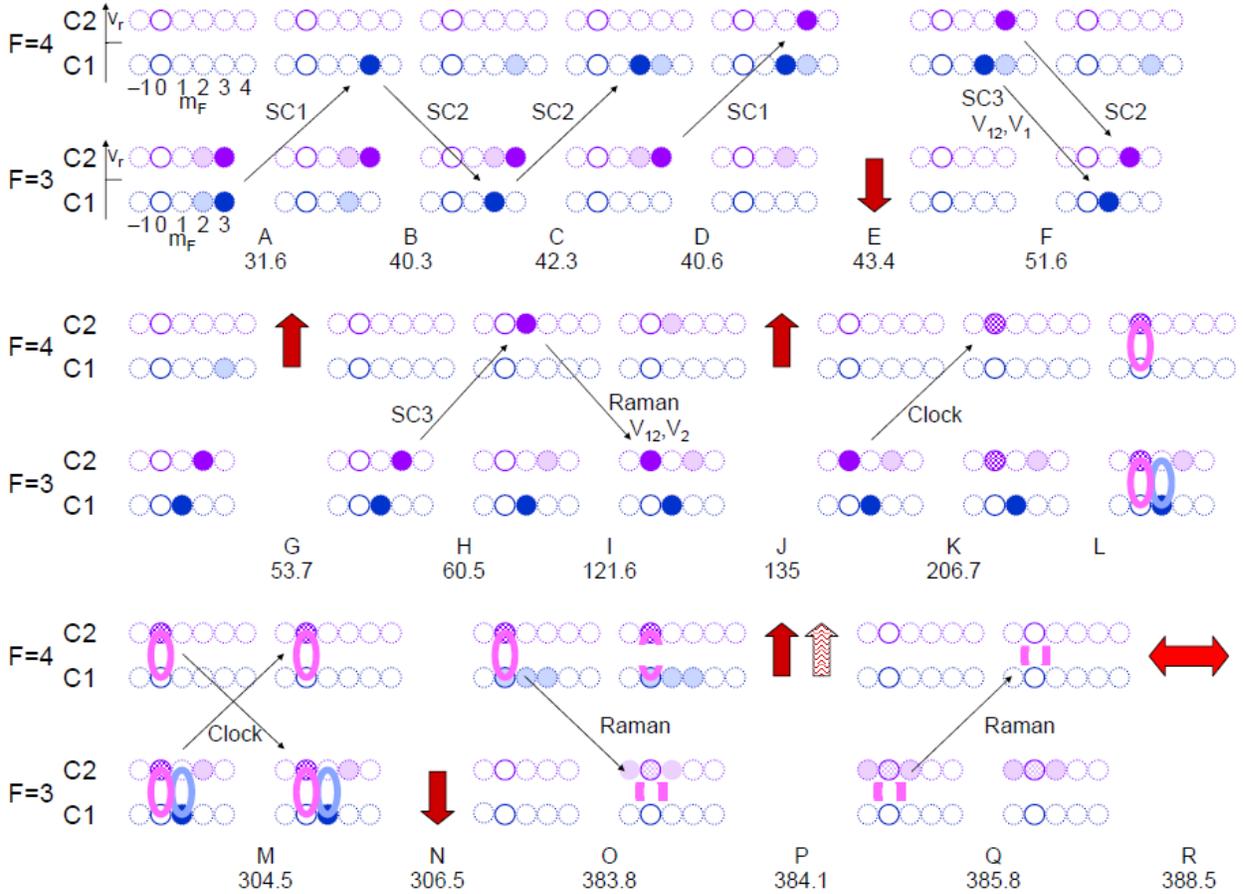

Fig. S2. Experimental sequence for 33.3 μK and C1 target atoms in $|31\rangle$. Each filled circle represents a cloud of atoms in the F=4 (top) or F=3 (bottom) hyperfine manifold, in a particular $m_F$ state. Time proceeds from A to R, with the time specified in ms after the lattice launch of cloud 1. The black arrows denote microwave or velocity-selective Raman transitions between $|3, m_F\rangle$ and $|4, m_F\rangle$ states, red arrows represent clearing laser pulses, and the ellipses, or parts thereof, scattered atoms. The lightly colored $m_F$ clouds are examples of small, unintended populations. To prepare $|3, m_F=2,3\rangle$ targets, the state preparation cavity pulses B and C are inhibited and the SC3 pulse F transfers the atoms to the target state, $|33\rangle$ or $|32\rangle$. The times are slightly different for other collision energies but the sequence is the same, except to prepare $|44\rangle$ for 12.2 μK. To prepare $|44\rangle$, pulses B, C, SC3 of F, G, J, N, and O are inhibited, and an extra F=3 clearing pulse is inserted between H and I.

After the state preparation, both clouds pass through the clock cavity, putting the atoms in C2 in a coherent superposition of $|30\rangle$ and $|40\rangle$ at step K in Fig. S2. Near the apogee at L, the two clouds collide and a small fraction of the atoms in the clouds scatter, forming an expanding spherical shell of atoms, whose coherence is shifted by the difference of their quantum scattering phase shifts. When the scattered shell is centered in the clock cavity, shortly before C2 is at the center of the cavity, the second π/2 clock pulse converts the phase of the scattered atoms' coherence into a population difference between $|30\rangle$ and $|40\rangle$ (M). Immediately after this pulse, an F=3 clearing pulse removes the scattered and the C2 unscattered atoms in $|30\rangle$, and the C1 target atoms. To detect the scattered atom's Ramsey fringe, we velocity-selectively transfer (O) atoms that scattered around 90° from $|40\rangle$ to $|30\rangle$, by detuning the Raman pulse to select the mean velocity of the two clouds. An F=4 clearing pulse then removes F=4 atoms at other velocities, including ≈99.9% of the unscattered C2 atoms. A depumping pulse tuned to F=4 → F'=4 ensures that no atoms remain in F=4. In step Q, another velocity selective Raman probe pulse transfers the same group of scattered atoms from $|30\rangle$ back to $|40\rangle$. Finally, a lens focuses laser-induced fluorescence, from a laser beam tuned to F=4 → F'=5, onto a photodiode. This laser beam is apertured by a 1 cm high horizontal slit to reduce the background from unscattered atoms.

In addition to detecting the phase of scattered atoms, we also measure the phase of the unscattered atoms (Fig. 2(b)). Here, we delay the second clock pulse until C2 is centered in the clock cavity. We apply the same sequence of clearing and Raman pulses in Fig. S2, with the Raman transitions tuned to resonance for the unscattered C2 atoms at O and Q. We also delay the detection pulse until C2 falls into the center of the masked detection beam.

To measure and remove backgrounds, we use a pump-probe technique where we inhibit either C1 or C2, or both [12]. The essential point is that scattered atoms can only be present if C1 and C2 collide. We remove C1 by inhibiting the microwave pulses B, C, and F, allowing the F=4 clearing pulse to remove C1 atoms. C2 is removed by tuning the first Raman pulse I off resonance by 1 MHz so that the F=4 clearing at J removes >99% of C2. From these configurations, we get the scattered atom Ramsey fringe in Fig. 2b, $V_s = V_{12} - V_1 - V_2 + V_b$, where both clouds give a Ramsey fringe $V_{12}$, just C2 gives $V_2$, only C1 gives $V_1$, and with both clouds cleared, $V_b$.

At each $E_c$ and B, we scan the clock frequency with 50 steps across the central 5 Ramsey fringes. There are 13 fountain launch cycles at each frequency, each lasting ≈1s. Six of these are of $V_{12}$ and $V_1$ for C1 prepared in $|3(1,2,3)\rangle$. Two are $V_2$ and $V_b$. For four, we rapidly reverse the magnetic field after the lattice launch to drive a Majorana transition to transfer C1 to $|3,-3\rangle$ and measure $V_{12}$, $V_1$, $V_2$ and $V_b$, since these backgrounds may be different. Finally, we probe the unscattered atoms to monitor the oscillator's frequency and magnetic field. To remove drifts in backgrounds, B, and our quartz oscillator's frequency, we reverse the order of these 13 measurements with each frequency step and reverse the frequency sweep direction for each successive scan of 50 frequency steps. We iteratively fit the Ramsey fringes to $\alpha\cos^2(\pi(f-f_0)/2\Delta\nu-\Phi)+\beta$. First we constrain $\Phi$ to be zero and fit the unscattered fringes, where the linewidth and central frequency $f_0$ are free parameters. We then fit the Ramsey fringes of the scattered atoms with $f_0$ fixed, first with $\Delta\nu$ free. We repeat the fit with $\Delta\nu$ fixed to the average value for the entire data set, yielding the scattered phase difference $\Phi$. At each magnetic field B, we average 3 scans of 50 steps, and sweep B between 25 and 1800 mG 6 times, for a total of 18 scans of 50 steps for each B. The phase shift $\Phi$ is a weighted average of N=18 measurements, where the weights come from the fit uncertainties $\sigma_i$ from each 50 step scan, after the backgrounds are subtracted. The error bars are $([\Sigma\sigma_i^{-2}(\Phi_i-\overline{\Phi})^2]/[(N-1)\Sigma\sigma_i^{-2}])^{1/2}$.

**Systematic errors:**

In this section we place bounds on several systematic errors. Because the state preparation in each fountain cycle always occurs with the same applied magnetic field, before the field is switched to the scanned field value B, systematic errors cannot produce the features in Figs. 2 and 3. Here, for completeness, we place limits on any phase offsets and on how much the offsets could change as the s-wave phase shifts vary significantly through scattering resonances.

Pump-probe techniques extract the change in the observed signal due to the presence of both the pump and probe. Systematic errors in the background subtraction occur when inhibiting one cloud produces unintended changes in the observed signal due to the other cloud. Here, because our Ramsey fringe measurements depend on the microwave phase difference between the two clock cavity passages, we are highly immune to several backgrounds, only being sensitive to atoms that occupy the clock states during the Ramsey time (steps K-M). As above,

$$V_1 = \Delta_1\left(-\tfrac{v_r}{2}\right)N_1 + V_b$$
$$V_2 = \int N_2 W_S(v_r)W_P(v_r)dv_r + V_b$$
$$\approx N_2 W_{sc}(0) + N_2 W_{un}\left(\tfrac{v_r}{2}\right) + V_b$$
$$V_{12} = (1-\varepsilon_2)[N_1 N_2 \zeta_{sc} + N_2 W_{sc}(0) + N_2 W_{un}\left(\tfrac{v_r}{2}\right)] + \Delta_1\left(-\tfrac{v_r}{2}\right)N_1 + V_b - \varepsilon_1 \Delta_1\left(-\tfrac{v_r}{2}\right)N_1$$

The background $V_b$ comes from incomplete clearing, mostly of C2, because its atoms are intentionally driven to m=0. With C2 inhibited, $V_1$ has a contribution from $V_b$ and any potential fraction $\Delta_1(-v_r/2)$ of C1 atoms with velocities near $-v_r/2$ that end up in m=0 before the clock pulse (K). For example, with $|31\rangle$ target atoms, spontaneous emission from the Raman pulse (I) will optically pump some atoms to $|31\rangle$ and produce a Ramsey fringe. However, the velocity-selective Raman probe pulses (O & Q) suppress it to negligible levels, and it is even more suppressed for other targets. With C1 inhibited, $V_2$ is the sum of $V_b$ and an integral over the probe's velocity selection probability $W_P(v)$ and the selected velocity distribution $W_S(v)$. The integral can be broken into two contributions, the wing of the probe exciting a small fraction of the large number of C2 atoms at $v_r/2$, and a small number of v=0 selected atoms that are resonant with the probe.

With both clouds enabled, $V_{12}$ has the contributions of $V_1$ and $V_2$, and from scattered atoms, which are proportional to $N_1 N_2$. Potential systematic background errors due to the inhibition of C1 and C2 are described by $\varepsilon_2$ and $\varepsilon_1$. As an example, the SC3 pulse (F) has some leakage into SC2. For $|32\rangle$ targets, the $|43\rangle \rightarrow |32\rangle$ SC3 frequency sweep for C1 will also drive this transition for C2 in SC2. This reduces the number of C2 atoms, both scattered and unscattered, only when C1 and C2 are both enabled ($V_{12}$). If C2 is inhibited, clearing pulse J effectively removes C2. A trivial effect is less scattering ($\zeta_{sc}$) and the important contribution is $\varepsilon_2$ [$N_2 W_S(0)W_P(0) \Delta v_P + N_2 W_S(v_r/2)W_P(v_r/2) \Delta v_{C2}$], which gives an incorrect subtraction of the C2 background of unscattered atoms. This always occurs for $|32\rangle$ targets, and for $|3,3(1)\rangle$ targets when the SC3 pulse is before(after) the SC2 pulse, which is the case for $E_c<(>)$ 38 μK. Measurements suggest the leakage could reduce the number of C2 atoms by 7%. In our worst case, $V_2$ has about twice the amplitude of $V_s$. With $\Phi=1$, the systematic error could be of order 0.1 rad, being smaller for $\Phi$ near 0. Comparing the data above or below 38 μK for $|33\rangle$ and $|31\rangle$ shows no clear systematic deviation. Since the $|3,-3\rangle$ preparation is unaffected by leakage, we can also compare $\Phi$ for $|33\rangle$ and $|3,-3\rangle$ targets, which must be identical as B→0. These, and a number of configurations of the experiment, including inhibiting pulse (A) to clear C1 with (F), showed no differences of order 0.03 rad. In future precision measurements, these and other sources of systematic errors, including those below, can and will need to be evaluated accurately.

Another systematic error comes from scattering events that happen between the Raman selection in step (I) and the first Ramsey pulse (K). They produce a scattered atom halo with a $\Phi=0$ Ramsey fringe. Their scattering cross section will be different as well since, during this time, C2 is in $|30\rangle$, instead of $|30\rangle$ and $|40\rangle$ after the first Ramsey pulse. Because the clouds do not substantially overlap until relatively close to (I), this is estimated to be less than 20% of the total observed scattered atom signal. It could reduce all phase shifts by as much as 15%.

Either cloud, C1 or C2, can also absorb spontaneously emitted photons while clearing the other cloud from the fountain (G, J, & N), which also yields an incorrect subtraction of a background Ramsey fringe. For G (J) clearing with light tuned to F=4 to F'=5, C2 (C1) is in F=3 and so the reradiated photon when C1(C2) hyperfine pumps to F=3 will be detuned by 250 MHz from the F=3 to F'=4 transition, giving a negligible absorption. Step N is F=3 clearing and, while the hyperfine depumping rates are higher, the absorption of reradiated photons is still negligible.

Similar to pulse F above, when the velocity-selective Raman pulse I is detuned to inhibit C2, it no longer selects a small fraction of the wing of C1 when C1 is a $|31\rangle$ target. Therefore, it also contributes an incorrectly subtracted background Ramsey fringe. However, because such a small fraction of C1 atoms normally end up in $|30\rangle$ before (K), this Ramsey fringe is extremely small. We note that we detune Raman pulse I, instead of inhibiting it, so that the Ramsey Fringe background due to any spontaneous emission from this Raman pulse is properly subtracted.

*Systematic errors in the Collision Energy:* We change the collision energy of the atoms by varying the time between the launches of the two clouds and adjusting independently the lattice launch-velocity of each cloud. The mean collision energy is $\bar{E}=m/4\ v_r^2 + k_B/4\ (T_{1x}+ T_{2x}+ T_{1y}+ T_{2y}+ T_{1z}+ T_{2z})$, where $T_{iv}$ is the Gaussian velocity width of C1 or C2 along the $v^{th}$ axis. The relative velocities of C1 and C2 range from $v_r$=5 cm/s and 11 cm/s, giving $\bar{E}$=(12.2, 16.9, 21.5, 33.3, 43.9, 49.2) µK. The velocity width of the atoms in the vertical z direction contributes most to the rms spread in collision energies, $\delta E^2 = m\ k_B/4\ v_r^2\ (T_{1z} + T_{2z}) + k_B^2/8[(T_{1x}+ T_{2x})^2 + (T_{1y}+ T_{2y})^2 + (T_{1z} + T_{2z})^2]$, giving ±(3.2, 4.1, 4.7, 6.9, 6.8, 8.9) µK. We monitor the velocity distributions in all three dimensions for both clouds with velocity-selective Raman transitions. Average velocity widths were $T_{1(2)z}$= 790(390) nK, $T_{1(2)x}$= 590(570) nK, and $T_{1(2)y}$= 1,390(1270) nK, stable to ±100 nK. We measure the relative velocities of the two clouds with velocity-selective Raman transitions and time-of-flight, detecting each cloud on the way up and down in the fountain.

Within this range of collision energies, our measurements may have a bias towards higher or lower energies. For example, the expanding spherical halo of scattered atoms must pass through the 18 mm diameter aperture of the clock cavity. Atoms that scatter early, or that have a high collision energy, will be preferentially cut by this aperture. This can be more serious for a higher fountain, and for higher collision energies, but we keep constant the spatial separation of C1 and C2 at detection, and hence the size of the halo. For collision energies below 30 µK, the fountain height was therefore taller. Overall we estimate that <10% of the scattered atoms are removed at all collision energies.

Limiting the fountain height shortens the time between the Ramsey interrogation pulses. At sufficiently short times, the two clouds do not completely pass through one another and the clouds may already begin colliding before the first Ramsey pulse. These collisions have an above average collision energy and do not contribute a Ramsey fringe, thus lowering the mean collision energy we observe. Collisions that would occur after the second Ramsey pulse are eliminated by the clearing pulse (N) in Fig. S2, thereby increasing the observed mean collision energy. We optimized the cloud overlap halfway through the interrogation time for $E_c$>20 µK, which minimizes this systematic energy shift. For $E_c<20$ µK the maximal overlap between clouds was closer to the second Ramsey pulse, giving an uncorrected bias to higher collision energies.

*Magnetic fields and gradients:* We apply a vertical ($\hat{z}$) magnetic field during the clock interrogation time with 5 independently adjustable coils surrounding the fountain. We change the currents of these coils ≈30 ms before the first Ramsey pulse, allowing for eddy currents to decay, until just after the second Ramsey pulse. The launch, state preparation, and detection are always done at the same magnetic field as we scan the pulsed field magnitude between 0.025 and 1.8 G.

We use the quadratic Zeeman shift of the unscattered atoms in C2 to measure the magnetic field. We first use a 50-100 ms Rabi pulse at the apogee and then 2-3 Ramsey fringe scans with increasing interrogation times, yielding the mean $B^2$ for C2. For small B, we also use the $|31\rangle \to |41\rangle$ field sensitive transition. This determines B to ±0.6 mG at B=25mG, decreasing to typically 0.05 mG above 0.6G. For B→0, we see a residual field of 16.3 mG. We also measure the magnetic field with a magnetometer ≈10 cm above the clock cavity, and add small currents to a set of coils to remove drifts of the vertical component of the magnetic field. The servo error fluctuates with a typical standard deviation of <50 µG.

Magnetic field gradients can produce a systematic phase error because the scattered atoms follow different trajectories than the unscattered atoms, which provide the phase reference. We shim the applied magnetic field so that the magnitude of the magnetic field B is uniform for C2 between the two clock pulses. If there are transverse gradients or curvatures of B, atoms following different trajectories will experience a different field, B+δB and therefore an additional quadratic Zeeman frequency shift 2 B δB T (427 Hz G$^{-2}$). This frequency shift produces a phase shift of 4πB δB T (427 mrad G$^{-2}$), where T is the time between Ramsey pulses.

Thus, a spatial magnetic field variation that does not change with B gives a phase shift that is linearly proportional to B. The magnetic field variation may also vary with the applied B, for example due to small gradients from the coils that make B, or variations from magnetic susceptibilities or long-lived eddy currents. Such variations produce phase shifts that increase quadratically with B.

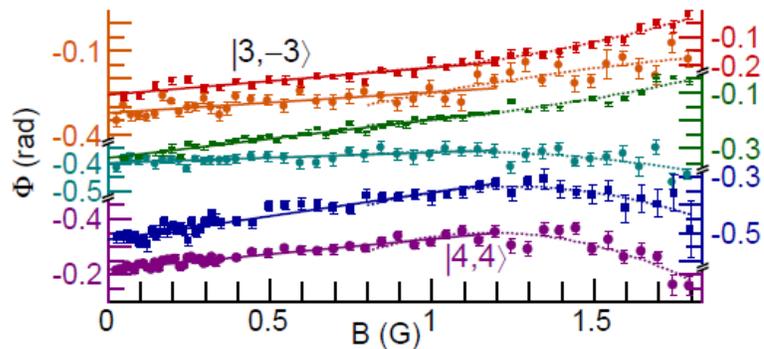

**Fig. S3**. Measured phase Φ for $|3,-3\rangle$ targets, and $|44\rangle$ for $E_c$=12.2 µK. We fit a line to the data between 0 and 1.2 G, and a quadratic from 0.8 G to 1.8 G, to correct for magnetic field variations in the fountain clock. At 1.8 G, 0.1 radians corresponds to a 38 µG difference in magnetic field for a 1.8 Hz Ramsey fringe linewidth.

To measure the magnetic field variations, we scatter clock state atoms off $|3,-3\rangle$ target atoms. The scattered clock atoms follow the same trajectories, independent of the target state. For $|3,-3\rangle$, Feshbach resonances are neither expected nor observed. We fit the measured phase versus B for $|3,-3\rangle$ at each collision energy to a line between 0 and 1.2 G, and a quadratic from 0.8 G to 1.8 G (see Fig. S3). To correct the measured phase for other target states, Figs. 2, 3 & S4, we use the linear fit up to 1 G and the quadratic from 1 to 1.8 G, requiring the correction to be 0 at B=0. After correcting for these magnetic field variations, the measured phases for each $|3,(1,2,3)\rangle$ target approach a constant at high B.

**Scaled Phase Shifts:**

Figure S4 shows the measured phase shifts $\Phi$ for $|33\rangle$ and $|31\rangle$ targets, where B is scaled by $E_c$ and the amplitude by the spread of collision energies $\delta E$, as in Fig. 2(c). The data are shifted vertically to match the 33.3 μK data.

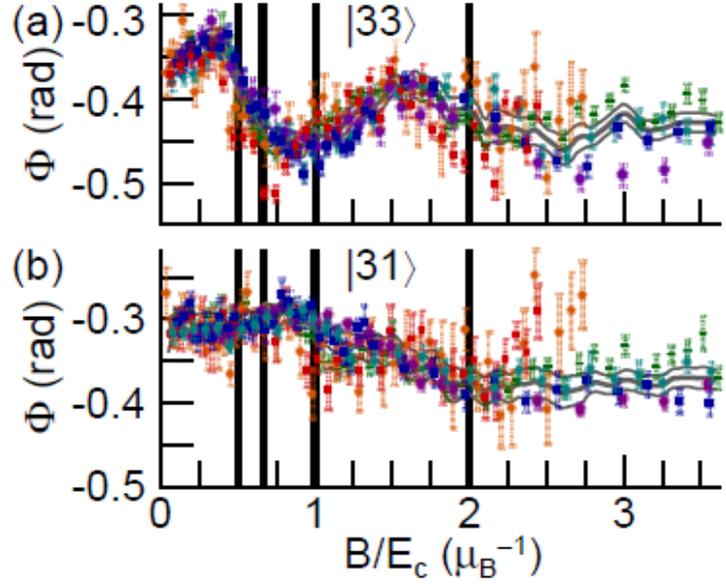

**Fig. S4**. Phase shift $\Phi$ for $|33\rangle$ (a) and $|31\rangle$ targets (b) versus B, scaled by $E_c$ and $\Phi$ scaled by the spread of collision energies $\delta E$ as in Fig. 2(c). The weighted average of $\Phi$ versus B gives the reference shape for all energies (grey curves) in Figs. 3(b) and (d).